# Transferring MBE-grown topological insulator films to arbitrary substrates and Metal-insulator transition via Dirac gap

Namrata Bansal[*], Myung Rae Cho[#], Matthew Brahlek[◊], Nikesh Koirala[◊], Yoichi Horibe[◊,‡], Jing Chen[◊], Weida Wu[◊], Yun Daniel Park[#], Seongshik Oh[◊,⊥]

[*] Department of Electrical and Computer Engineering, Rutgers, the State University of New Jersey, Piscataway, NJ 08854, USA

[#] Department of Physics and Center for Subwavelength Optics, Seoul National University, Seoul, 151-747, South Korea

[◊] Department of Physics and Astronomy, and Rutgers Center for Emergent Materials, Rutgers, the State University of New Jersey, Piscataway, NJ 08854, USA

[‡] Department of Materials Science and Engineering, Kyushu Institute of Technology, Kitakyushu, Fukuoka, 804-8550, Japan

[⊥] Institute for Advanced Materials, Devices and Nanotechnology, Rutgers, the State University of New Jersey, Piscataway, NJ 08854, USA

**ABSTRACT**

Mechanical exfoliation of bulk crystals has been widely used to obtain thin topological insulator (TI) flakes for device fabrication. However, such a process produces only micro-sized flakes that are highly irregular in shape and thickness. In this work, we developed a process to transfer the entire area of TI $Bi_2Se_3$ thin films grown epitaxially on $Al_2O_3$ and $SiO_2$ to arbitrary substrates, maintaining their pristine morphology and crystallinity. Transport measurements show that these transferred films have lower carrier concentrations and comparable or higher mobilities than before the transfer. Furthermore, using this process we demonstrated a clear metal-insulator transition in an ultrathin $Bi_2Se_3$ film by gate-tuning its Fermi level into the hybridization gap formed at the Dirac point. The ability to transfer large area TI films to any substrate will facilitate fabrication of TI heterostructure devices, which will help explore exotic phenomena such as Majorana fermions and topological magnetoelectricity.

Topological Insulators (TI) are nonmagnetic insulators that have a gapped bulk band but protected metallic surface states as a consequence of spin-orbit interactions and time-reversal symmetry.[1-4] Band structure topology of these materials coupled with the spin-momentum locking mechanism prevents these metallic surface states from backscattering and localization.[5-9] $Bi_2Se_3$ is one of the most extensively studied TI materials as it has the simplest surface state structure and a relatively large band gap of 0.3 eV.[10] $Bi_2Se_3$ has a rhombohedral structure with a sequence of Se-Bi-Se-Bi-Se forming a quintuple layer (QL, 1 QL $\approx$ 1 nm) along the c-axis direction. This layered structure is held together strongly within the QL but only weakly between layers by van der Waals forces.[11-12] Significant progress has been made in understanding the properties of TIs through angle resolved photoemission spectroscopy (ARPES),[8,13-15] scanning tunneling microscopy[7] and transport measurements.[16-25] Recently, there is a growing interest in combining TIs with superconductors, (anti-)ferromagnets etc. in search of quantum anomalous Hall effect,[26-27] Majorana fermions[28-29] or other magnetoelectric effects.[30-31]

The weak inter-layer bonding enables mechanical exfoliation of thin $Bi_2Se_3$ flakes from bulk crystals and such flakes have been extensively used to fabricate TI devices.[16-20,32-33] However, the process of mechanically cleaving bulk crystals results in flakes of limited lateral sizes and ill-defined thicknesses, and the yield of obtaining ultrathin flakes ($\leq$ 10 nm) is very low.[20,32,34-35] On the other hand, techniques to transfer large-area 2D layered materials are well established in the field of graphene. A polymeric supporting layer has been successfully demonstrated as a transferring template for various 2D materials such as graphene,[36-37] hexagonal boron nitride,[38] molybdenum disulphide[39] etc. However, such a process has not been developed for TI structures yet. Herein, we demonstrate that large-area $Bi_2Se_3$ films grown by molecular beam epitaxy (MBE) on $Al_2O_3(0001)$ and amorphous $SiO_2$ (a-$SiO_2$) substrates can be transferred onto various substrates, including a common transparency sheet. This transfer process opens a new door for TI research in that MBE allows not only precise control in film thickness but also formation of novel heterostructures that cannot be achieved with other synthesis methods.

High-quality $Bi_2Se_3$ films were grown on 1 cm $\times$ 1 cm $Al_2O_3(0001)$ substrates[24] and 3-inch oxidized Si wafers[40] using a custom-designed SVTA MOS-V-2 MBE system. The transfer of these films to various substrates was carried out using a poly(methyl methacrylate) (PMMA) layer as a transfer support; a single coat of PMMA provides enough mechanical strength to the $Bi_2Se_3$ film to prevent any cracking during the transfer process. Fig. 1a shows the schematic for this process. After PMMA (A-6) is spin-coated and baked at 100 $^o$C on top of $Bi_2Se_3$ film, the sample is floated on buffered oxide etch (BOE) or potassium hydroxide (KOH) solution. The solution acts on the interface between the film and the substrate and the peeling of PMMA/Bi$_2$Se$_3$ layer thus initiates from outer edge of the sample. Then, probably by capillary wet etch mechanism, the pealing gradually spreads out (Fig. 1b,c). This results in a free-standing PMMA/Bi$_2$Se$_3$ film that can be transferred to any desired substrate after rinsing off any excess BOE/KOH using de-ionized water (Fig. 1d,e). After drying the film on the new substrate in ambient air, the PMMA layer can be easily removed by acetone. The detachment of PMMA/Bi$_2$Se$_3$ layer from the underlying substrate was much faster for films on a-SiO$_2$ than on Al$_2$O$_3$(0001). This process allows transfer of very large area Bi$_2$Se$_3$ films of any thickness to any substrate for multiple applications ranging from back-gating to flexible electronics; Fig. 1f,g show a 30 QL Bi$_2$Se$_3$ film of area ~3 cm × 3 cm transferred to a common transparency sheet.

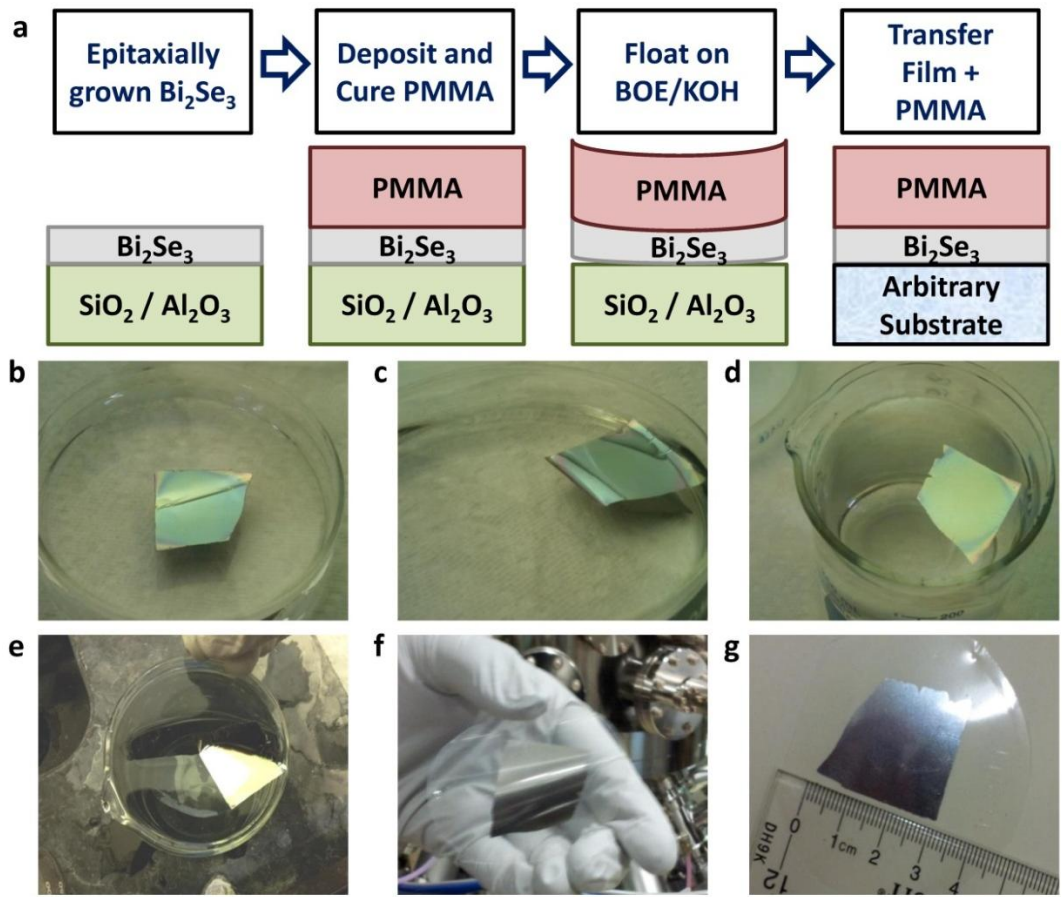

**Figure 1. Process for transfer of Bi$_2$Se$_3$ thin films. a.** Schematic diagram of the PMMA-based transfer method of Bi$_2$Se$_3$ films grown on a-SiO$_2$ or Al$_2$O$_3$(0001) to any arbitrary substrate. **b-e.** Photographs of a 30 QL thick Bi$_2$Se$_3$ film grown on a-SiO$_2$ transferred to a common transparency sheet. (**b,c**) PMMA spun on the sample provides enough buoyancy for floating the sample on BOE; as the peeling of the film progresses, the substrate sinks down. (**d**) The PMMA/Bi$_2$Se$_3$ film is then floated in multiple water baths to remove any excess chemicals. (**e**) The PMMA/Bi$_2$Se$_3$ film can be fished out of water using target substrate, a piece of common transparency sheet in this case. (**f**) Film transferred on a flexible substrate allows bending of the film without any cracks visible under an optical microscope. (**g**) Photographs of a ~3 cm × 3 cm Bi$_2$Se$_3$ film transferred from a-SiO$_2$ to a transparency sheet.

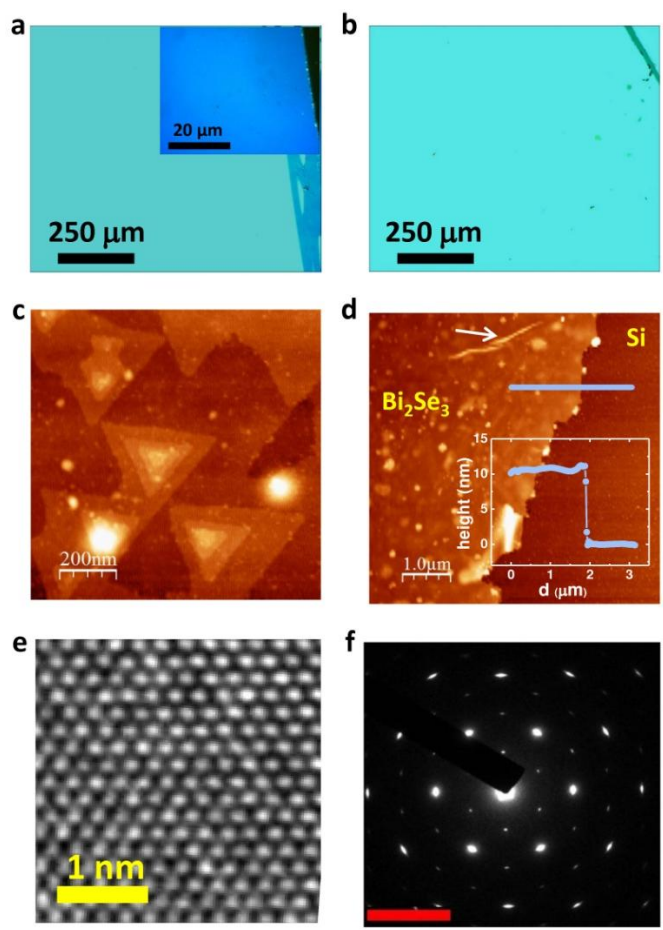

**Figure 2. Characterization of the transferred film. a-b.** Optical images of a 10 QL $Bi_2Se_3$ film grown on $Al_2O_3$(0001) transferred to (**a**) $Al_2O_3$(0001) and (**b**) Si(111). The inset in (a) highlights the edge of the transferred film on $Al_2O_3$ substrate. The films are intentionally scratched on the sides to get a clear contrast. **c.** AFM image of the 10 QL film transferred to Si(111), exhibiting the pristine morphology of the transferred film. This image confirms that the chemicals used in the process do not attack the film. **d.** A large area AFM image (5 × 5 $\mu m^2$) showing the edge of the transferred film; the inset shows the height profile of the film along the line drawn on the image. The measured height (~10 nm) is consistent with the original thickness of 10 QL. The feature pointed by the arrow is probably a micro-wrinkle in the transferred film along the edge. **e. Plan view** TEM image obtained from a 6 QL film grown on $Al_2O_3$(0001) and transferred to a carbon foil and **f.** corresponding SAED pattern acquired from the same sample. The scale bar is 5 $nm^{-1}$.

Figure 2 provides the morphological and structural characteristics of the transferred films in optical microscopy (OM), atomic force microscopy (AFM), and transmission electron microscopy (TEM) images. TEM characterization was performed on a 6 QL $Bi_2Se_3$ film grown on $Al_2O_3$(0001) and transferred to a carbon foil on a TEM copper grid. A JEOL-2010F transmission electron microscope (TEM) equipped with a 14-bit charge coupled device (CCD) array detector was used to take the electron diffraction patterns and high-resolution images. Fig. 2a,b show optical images of a 10 QL thick $Bi_2Se_3$ film grown on $Al_2O_3$(0001) and transferred to $Al_2O_3$(0001) and Si(111) substrates; part of the film is intentionally scratched to get a clear contrast. The excellent continuity of the film over a few millimeter length scale with uniform color contrast and absence of any wrinkles or bubbles in the images indicate the thickness uniformity and efficiency of the transfer process. Atomically flat film with long triangular terraces are observed in the AFM image (Fig. 2c) of a 10 QL-thick $Bi_2Se_3$ film transferred to Si(111); there is no surface modification in the transferred film and the morphology looks similar to that of a pristine $Bi_2Se_3$ film. A large area, $5 \times 5$ μm$^2$, AFM scan is shown in Fig. 2d; the inset shows the height profile of the transferred film along the marked line. The large area scan shows one anomalous wavy feature, as pointed out by the arrow, which is probably due to a micro-wrinkle formed during the drying of the transferred film. The hexagonal atomic lattice structure seen in the TEM image of a 6 QL film (Fig. 2e) and the corresponding hexagonally symmetric pattern of the selected area electron diffraction (SAED) (Fig. 2f) also confirm the excellent quality of the transferred film.

$Bi_2Se_3$ films on $Al_2O_3$ substrates were cut into four (5 mm × 5 mm) pieces after spin-coating and curing PMMA on top; three of them were transferred onto $Al_2O_3$(0001), Si(111) and a-$SiO_2$ substrates, and the last piece was kept 'as-is' to compare the transport properties before and after transfer. After the PMMA was removed from each sample, transport measurements were carried out with the van der Pauw geometry in an AMI superconducting magnet with a base temperature of 1.5 K and a maximum field of 9 T. It was verified that PMMA curing, without transfer, on a pristine sample caused no significant degrading in the transport properties. The nonlinearity in the Hall resistance, $R_{xy}$, of the transferred films of varying thicknesses, Fig. 3a, implies that multiple channels contribute to the conduction; thus, only the total carrier concentration obtained from high B-field slope was considered. Figures 3b-d compare the transport properties of 10 QL (Fig. 3b) and 32 QL (Fig. 3c) $Bi_2Se_3$ films grown on $Al_2O_3$(0001) and a 100 QL (Fig. 3d) $Bi_2Se_3$ film grown on a-$SiO_2$ before (as-is) and after transferred onto different substrates. Due to the environmental doping effect[41], exposure of $Bi_2Se_3$ films to air tends to increase the n-type carrier densities of the films with time. As these samples were left exposed to air for an indeterminable time, the 'as-is' samples have higher sheet carrier density and lower mobility compared to pristine samples[24,40]. However, it is interesting to note that despite the aging effect in $Bi_2Se_3$ films, the transferred films have lower carrier concentrations than not only the 'as-is' and but also the pristine samples[24,40], especially in thinner films. This suggests that the transfer process is quite robust and is not affected much by aging of the samples. We recently reported that though the high quality $Bi_2Se_3$ films on $Al_2O_3$(0001) have very low bulk carrier densities, their surface charge defect density is quite substantial.[24,40] A plausible cause for the reduction in the carrier density after transfer is that the interface etching mechanism involved in the transferring process may either remove or neutralize the surface charge defects. This scenario is also supported by the observation that unlike the thinner samples, the change in the sheet carrier density of the thick 100 QL film grown on a-SiO$_2$ substrate during the transfer is relatively insignificant; due to comparatively large bulk contribution in the thicker films grown on a-SiO$_2$,[40] it is expected that the surface charge accounts for only a small fraction of the total sheet carrier density in these samples.

The change in mobility of the film after transfer also follows the above scenario. Mobility is a measure of scattering time and thus it is quite susceptible to disorders and interactions. Interestingly, Figure 3 shows that most of the transferred films in the thin regime had their mobilities enhanced rather than degraded after the transfer. This enhancement in mobility indicates that the transfer process improves the film quality by removing the surface charge defects; this is quite surprising considering all the chemicals involved during the process. To demonstrate the applicability of this method to large-scale flexible electronics, we also measured a 100 QL film transferred from a-SiO$_2$ to a common transparency sheet as shown in Fig. 3d. The comparatively low mobility (~850 cm$^2$/Vs) observed in this sample (labeled 'Trans.' in Fig. 3d) is probably due to the graininess of the transparency sheet causing unevenness in the overlaying Bi$_2$Se$_3$ layer; still, the carrier density is comparable to those transferred onto other highly polished substrates.

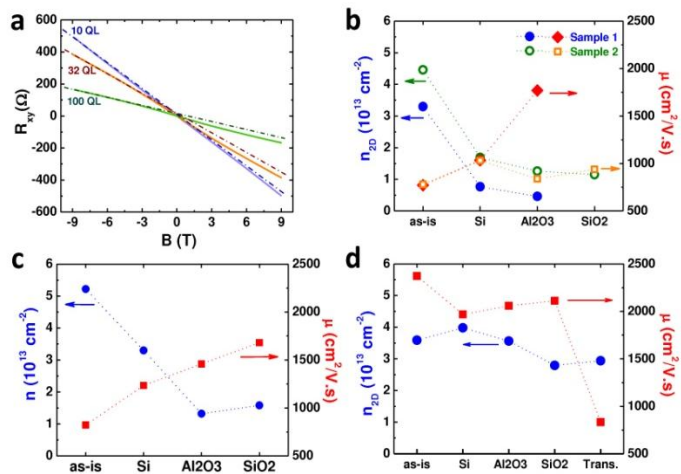

**Figure 3. Transport properties. a.** Hall resistance vs. B for 10 QL, 32 QL and 100 QL Bi$_2$Se$_3$ films transferred to a-SiO$_2$. The nonlinearity of the Hall resistance is indicative of multiple conducting channels; the dashed lines are straight guide lines, showing the level of non-linearity for each curve. All samples are measured at 1.5 K, and due to nonlinearity in the Hall resistance of these samples, high field (~9 T) slope was used to estimate the total sheet carrier densities. Comparison of sheet carrier densities and corresponding mobilities of **b.** 10 QL, **c.** 32 QL thick Bi$_2$Se$_3$ films grown on Al$_2$O$_3$(0001) and **d.** 100 QL Bi$_2$Se$_3$ film grown on a-SiO$_2$ transferred to different substrates, as labeled; the point 'Trans.' refers to the film transferred to a common transparency sheet, and the point labeled 'as-is' is the part of the film as grown on the initial substrate and not transferred.

For the films transferred onto a-SiO$_2$, the Hall bar device was fabricated by ion milling in argon plasma through a shadow mask; the doped Si substrate was used as a back gate. It has been previously reported that argon bombardment induces hysteric metallic states on the surface of STO.[47-48] So for films transferred onto STO, hand patterning instead of ion milling was used to make the Hall bar pattern, and silver paint was used to make the back contact. In both cases, the Hall bar width was 0.56 mm and the length measured between the mid points of the longitudinal voltage probes was 2.1 mm. The gating measurements were conducted using an electromagnet at a base temperature of 6 K and maximum magnetic field of 0.6 T. Fig. 4a shows a photograph of a Hall bar device with an 8 QL film grown on Al$_2$O$_3$(0001) and transferred to a 5 mm × 5 mm x 0.5 mm SrTiO$_3$ (STO) substrate; the substrate acts as a dielectric for back-gating purpose. The initial carrier density of ~8 × 10$^{12}$ cm$^{-2}$ implies that the Fermi level is low enough to be in the bulk band gap with no bulk contribution.[35] This value of carrier concentration is lower than that of films of similar thickness grown directly on STO.[42-43] Fig. 4b shows the sheet resistance, $R_{xx}$, the low B-field Hall coefficient, $R_H$, (top panel), and corresponding sheet carrier density ($n_{2D} = 1/(eR_H)$) and mobility (bottom panel) as a function of applied back-gate voltage, $V_G$. By negatively biasing the gate, n-type carriers are depleted from the film, which is reflected as an increase in $R_{xx}$ and decrease in $R_H$. Beyond $V_G$ = -23 V, corresponding to an electron density of 1.7 × 10$^{12}$ cm$^{-2}$, $R_H$ starts increasing and system crossovers from a pure electron regime to an electron-hole regime. $R_H$ changes sign completely near $V_G$ = -39 V; this can be understood as Fermi level moving below Dirac point for both top and bottom surfaces and the transition of the system to a pure hole regime. As more negative bias is applied, more p-type carriers are injected, leading to a continuous drop in $R_{xx}$. Similar behavior was observed for an 8 QL Bi$_2$Se$_3$ film grown on a-SiO$_2$ and transferred back to a-SiO$_2$, using the doped Si as the back gate (Fig. 4c). Again, by applying negative gate bias we were able to reduce the electron density from 8.8 × 10$^{12}$ cm$^{-2}$ at $V_G$ = 0 V to 1.5 × 10$^{12}$ cm$^{-2}$ at $V_G$ = -162 V. P-type conduction was observed for $V_G$ < -185 V accompanied with a maximum in longitudinal resistance, $R_{xx}$.

Finally, we demonstrate the electrostatic control of Fermi level through the hybridization gap opened at the Dirac point of a 4 QL thick Bi$_2$Se$_3$ film due to coupling of the top and bottom surfaces.[44] Regions of different thicknesses within a mechanically exfoliated Bi$_2$Se$_3$ crystal on a-SiO$_2$ have often been identified by color contrast of the film with respect to the substrate.[20,35] As observed by the uniformity in color in the optical image of a 4 QL film transferred from Al$_2$O$_3$(0001) to a-SiO$_2$ (Fig. 4d), large-size ultrathin films of uniform thickness can be easily transferred by this process. Another major advantage is that, so far, some of the highest quality films have been achieved on Al$_2$O$_3$(0001)[24,45] and this method provides an efficient way to transfer these high quality films to arbitrary substrates. This is best exemplified by the observation that the resistance of the 4 QL film grown on Al$_2$O$_3$(0001) and transferred onto STO shows metallic temperature dependence all the way down to the lowest measured temperatures (Fig. 4e). In such an ultrathin regime, only high mobility films show metallic temperature dependence,[24] and films with low mobilities exhibit insulating temperature dependence as observed in ultrathin films grown directly onto STO,[46] cleaved from bulk[35] or exfoliated from nano-ribbons,[34] even if their Fermi levels are far from the hybridization gap. Therefore, the very observation of metallic temperature dependence of the 4 QL sample both before and after the transfer further evidences the robustness of this transfer process.

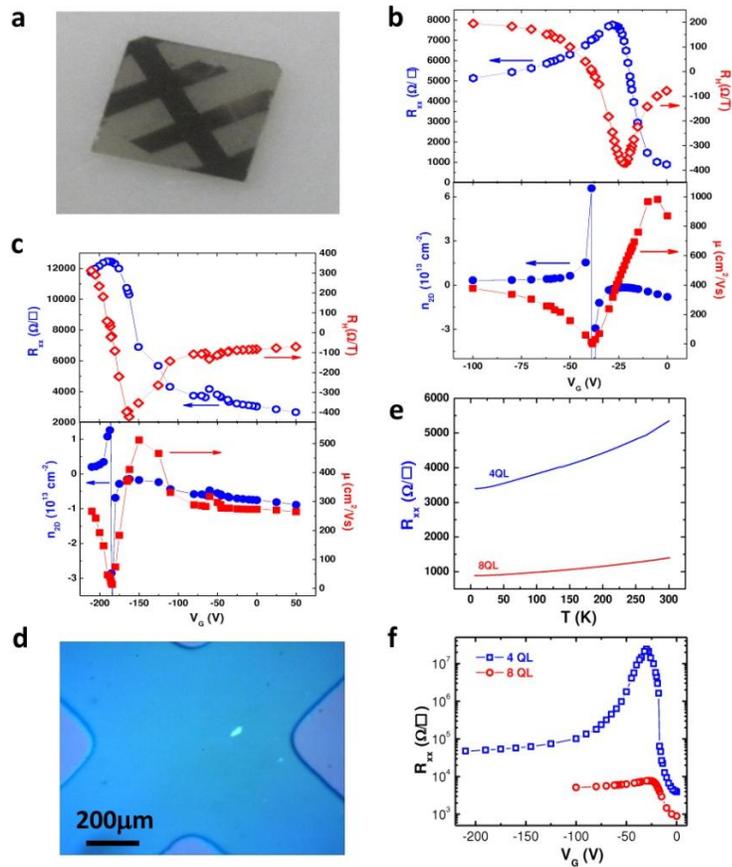

**Figure 4. Electrostatic back-gating of ultra-thin transferred films. a.** Photograph of an 8 QL $Bi_2Se_3$ film grown on $Al_2O_3(0001)$ and transferred to STO (5 mm × 5 mm); the Hall bar pattern was made manually through a metal shadow mask. **b,c.** Longitudinal resistance, Hall resistance (top panel), and corresponding sheet carrier densities and mobility (bottom panel) of (**b**) an 8 QL-thick film grown on $Al_2O_3(0001)$ and transferred to STO, and (**c**) an 8 QL-thick film grown on a-$SiO_2$ and transferred back to a-$SiO_2$. **d.** An optical image of a Hall bar device of a 4 QL $Bi_2Se_3$ film grown on $Al_2O_3(0001)$ and transferred to a-$SiO_2$; absence of any color change highlights the thickness uniformity of the ultra-thin transferred film. **e.** Resistance vs temperature for 8 QL and 4 QL $Bi_2Se_3$ films transferred onto STO; both show metallic temperature dependence. **f.** Gate-voltage dependent longitudinal sheet resistance of an 8 QL-thick film (red circles) and a 4 QL-thick film (blue squares) grown an $Al_2O_3$ and transferred to STO as a function of back-gate voltage. Although four-probe method was used for all other cases, two-probe method had to be used for the highly resistive state ($V_G < -18$ V) of the 4 QL film.

Figure 4f shows the gate voltage dependence of $R_{xx}$ for 4 and 8 QL thick films transferred from $Al_2O_3$ to STO. For the 8 QL film, $R_{xx}$ was measured in a four-probe configuration, as shown in Fig. 4a. The 4 QL film was metallic with $n_{2D}$ = 6.2 × $10^{12}$ $cm^{-2}$ and μ = 270 $cm^2$/Vs at $V_G$ = 0 V but it soon became highly insulating on applying negative gate bias; for $V_G$ < - 18 V, the sample became so insulating that we had to switch to the two probe method to measure the resistance in this regime, but we confirmed that the two point and the four point method provide consistent resistance values in the metallic regime. For both 4 and 8 QL films, the resistance increased as a negative gate voltage was applied, reached a maximum, and decreased with further increase in the magnitude of the gate voltage. This implies that the Dirac points were reached in both cases. However, a closer look reveals significant qualitative difference between the two. For the 8 QL film, the surface states remained metallic at all times with the resistance well below the quantum resistance ($h/e^2$) of 25.8 kΩ and the peak resistance (7.8 kΩ/sq.) was about ten times larger than the ungated value. On the other hand, even if the ungated 4 QL film was fully metallic with its resistance substantially lower than the quantum resistance, gating made the resistance cross the quantum resistance at $V_G$ ≈ -15 V and reach a maximum value of ~25 MΩ/sq, which is 1000 times larger than the quantum resistance, covering four orders of resistance modulation. In other words, while the 4 QL film is metallic when its Fermi level is far from the Dirac point, it becomes highly insulating as the Fermi level approaches the Dirac point.

The presence (absence) of the metal-insulator transition in the 4 QL (8 QL) film imply that the 8 QL film does have gapless Dirac states whereas the 4 QL sample has a gap at the Dirac point. According to the previous ARPES study, it is known that a gap is formed at the Dirac point for ultrathin TI films due to hybridization of the top and bottom surface states.[44] According to this observation, if the Fermi level is tuned in and out of the Dirac point, metal-insulator transition is obviously expected. However, despite numerous gating studies so far, no such observation has been made until now. There are multiple factors responsible for that. Sometimes, due to the difficulty in quality control in the ultrathin regime, ultrathin films exhibit extremely low mobilities; in such a case, the film remains insulating throughout the gating.[35] In other cases, the carrier density (surface or bulk or both) is so high that the Fermi level cannot be fully tuned into the gap; in that case, the film remains metallic throughout the gating. The very observation of a clear metal-insulator transition in the transferred 4 QL film implies that we have effectively overcome these barriers by combining high quality film growth with the effective film transfer method. This work also demonstrates that unlike the graphene system, it is possible to turn off all the conducting channels in properly engineered TI films, which is a highly desired property for device applications.

To summarize, we have shown that high-quality MBE-grown $Bi_2Se_3$ films of any thickness, as low as 4 nm, can be easily transferred to a whole range of desired substrates, including flexible transparencies.  Using this process, we demonstrated complete suppression of the metallic transport channel in an ultrathin film by tuning the Fermi level into the hybridization gap formed at the Dirac point.  The efficient back-gating effect also suggests that the substrate side of the transferred films is free of significant defects, which could work as traps for mobile carriers.  Considering that MBE allows atomic-level control of various heterostructures, transferring MBE-grown TI films to arbitrary templates will bring many new opportunities for novel TI devices.

## AUTHOR CONTRIBUTION

N.B., M.B. and N.K. grew $Bi_2Se_3$ films.  M.R.C. and Y.D.P. developed the transfer process.  N.B. transferred the films, fabricated devices and did the measurements.  Y.H. did the TEM study. J.C. and W.W. carried out the AFM measurement.  S.O. supervised the research.  N.B. and S.O. wrote the paper, with inputs from all other co-authors.

## AUTHOR INFORMATION

**Corresponding Author**

Email: ohsean@physics.rutgers.edu

## ACKNOWLEDGEMENTS

This work is supported by National Science Foundation (DMR-0845464 and DMR-0844807) and Office of Naval Research (N000140910749).  M.R.C. and Y.D.P are supported by the National Research Foundation of Korea grant, funded by the Korea government (MSIP) (No. 2013-030172, 2013-043649, 2008-0061906).